# Improving parent-austenite twinned grain reconstruction using electron backscatter diffraction in low carbon austenite

Ruth M. Birch[1]*, T. Ben Britton[1], W. J. Poole[1]

1. Department of Materials Engineering, University of British Columbia, Frank Forward Building, 309-6350 Stores Road, Vancouver, BC, Canada V6T 1Z4

*corresponding author: ruth.birch@ubc.ca

**Abstract**

Thermomechanical controlled processing (TMCP) is widely used to optimize the final properties of high strength low alloy (HSLA) steels, via microstructure engineering. The room temperature microstructures are influenced by the high temperature austenite phase, and the austenite microstructure is commonly accessed by reconstruction using electron backscatter diffraction (EBSD) data of the final microstructure. A challenge for reconstruction of the PAG microstructure and subsequent austenite grain size measurement is the presence of austenite-phase annealing twins, and we address challenge with a new 're-sort' algorithm. Our algorithm has been validated using the retained austenite regions (which were recovered via advanced pattern matching of EBSD patterns). We demonstrate that the re-sort algorithm improves the PAG reconstruction significantly, especially for the grain boundary network and correlation with other methods of grain size assessment and development of TMCP steels.

**Graphical Abstract**

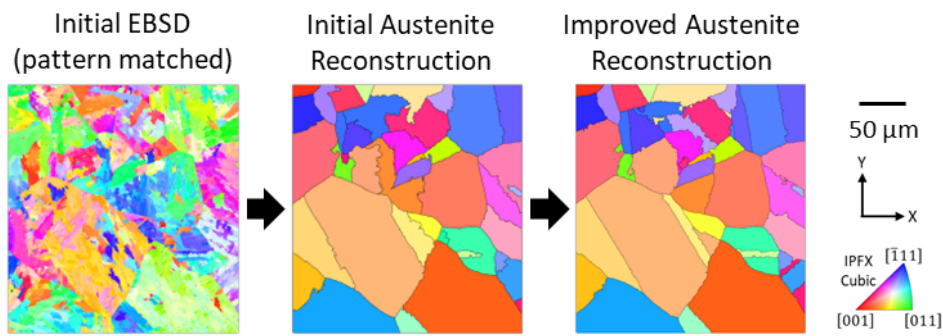

Thermomechanical controlled processing (TMCP) is widely used to control the microstructure and thereby the final properties of high strength low alloy (HSLA) steels. In these processing routes, the high temperature austenite grain structure influences the decomposition of austenite on the runout table after high temperature rolling; for example, grain boundaries are well known to be nucleation sites during decomposition for both ferritic and bainitic final microstructures. Effective refinement of the final microstructure is the key to an improved combination of strength, impact toughness and lower ductile-brittle transition temperatures. However, it is challenging to directly observe microstructures in the austenite phase field (above 900 °C for the current low-carbon steels of interest in this study), with the options being small scale *in-situ* EBSD [1], X-ray synchrotron experiments [2] or indirect methods such as correlating the attenuation of ultrasonic waves with austenite grain size during the thermal cycle [3].

Historically, prior austenite grain (PAG) boundaries have been revealed using chemical [4] or thermal [5] etching, or by reconstructing the PAG microstructure from the EBSD data [6–11] as will be discussed in this paper.

PAG reconstruction codes use the orientation data for the final microstructure and the orientation relationship (OR) between the high temperature austenite (face centred cubic, FCC) and low temperature body centred cubic (BCC) phases to reconstruct the austenite microstructure. The orientation relationship varies depending on composition and processing route, though the most commonly used orientation relationships are the Kurdjumov-Sachs (KS) or the Nishiyama-Wasserman (NW). These are similar but have a rotation about the closest packed plane, {111}, of 5.26° [12] .

A challenge for the reconstruction of the PAG microstructure and subsequent austenite grain size measurement is the presence of austenite-phase annealing twins [13]. These twins can potentially act as nucleation sites for austenite decomposition and they have also been shown to affect grain size measurements by Nyyssönen et al. [14]. Annealing twin boundaries form $\Sigma3^n$ interfaces, and the most common $\Sigma3$ interface has a misorientation of 60°<111>. The $\Sigma3$ interface has local atom-plane stacking of ABCBA, where the 'C' plane is common between both crystals, and if the C-plane is the interfacial habit plane then this called a coherent twin plane. The low energy of the coherent interface often means that the coherent $\Sigma3$ is flat in 3D (i.e. a long line in a 2D section), and so they can often easily be observed by eye in micrographs of FCC materials that contain annealing twins [15,16]. For PAG reconstruction, the orientation relationship of the twin-related austenite crystals results in potential child variants that have a common crystal whether they originate from either twin-related austenite crystal. In the case of the Kurdjumov-Sachs orientation relationship, an austenite grain and its twin would have 6 common child variants leading to a degree of ambiguity for the reconstruction [10]. The twin-related challenge is shown in Figure 1 where the two twinned-austenite grains have been rotated so that the shared [1 1 1] is out of plane, and the shared variants are indicated by the blue circles with red centres.

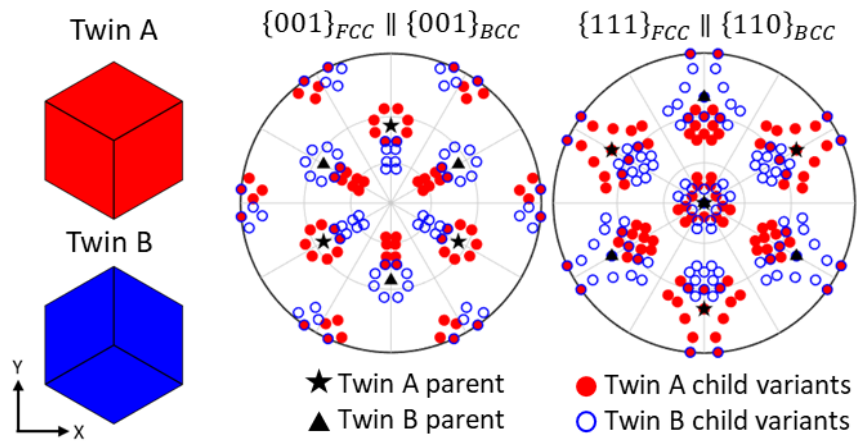

*Figure 1 – Example showing an example pair of twinned austenite (FCC) orientations and their child variants, as calculated using the Kurdjumov-Sachs orientation relationship. The pole figures are {001} in both FCC and BCC and {111}$_{FCC}$ overlaid with {110}$_{BCC}$ – black markers are used for the parent (FCC) phase whilst the twin variants are in red and blue, matching the parent unit cell colour for each twin austenite orientation.*

Accurate reconstruction of the parent austenite grain and its orientation can be identified if there are enough variants within the child microstructure, but the presence of common austenite-twin-related child variants near the austenite-twin interface can result in the incorrect positioning of the twin-boundary in the reconstructed micrograph. Typically, this can be observed by a careful (human-based) interpretation of the child and parent microstructure 2D maps, where the potential of a straight twin-interface is explored (for a 2D map) and twin-related variants could be re-attributed to the other twin-related parent austenite grain.

In practice, it would be ideal to include these corrections within the reconstruction code and there are some algorithms that have been developed to account for twinning [7,17–20]. Some of these have specific use cases, for example, Sun et al. [17] looked at twins in ultrafine grain prior austenite that have minimal child variants present. Other algorithms are computationally expensive and/or semi-automated, as in the case of neighbour-to-neighbour reconstruction approaches, with delayed decision making [18] or regional voting [20] to consider multiple PAG orientation options before the final decision is made.

Recent work featuring comparisons of PAG grain identification highlights the limitations when it comes consistent twin identification [1,21]. The method in Oxford Instruments AZtec crystal (which is based on work by Huang et al. [20] and includes regional voting) and MTEX variant graph [7] are both capable of consistently identifying the presence of twins within grains, but often with irregular boundaries [14], which may be sufficient for (merged) grain size analysis but not for local twin boundary related investigations.

In this manuscript, we present an alternative approach where twins are identified through a 're-sort' algorithm after (any) initial reconstruction of the PAG microstructure to improve the identification of twins in reconstructed PAG microstructures. The algorithm used here has been developed using MATLAB and the MTEX toolbox [22] (v 5.8.2), allowing the user to choose their preferred initial PAG reconstruction method. The code is available on Zenodo at https://doi.org/10.5281/zenodo.13380514.

To demonstrate this method, analysis was performed on a line pipe steel with the composition at Table 1. The sample was heated to 1300 °C at 50 °C/s with a hold at 1300 °C for 20 °s, cooled to 900 °C at 15 °C/s, cooled from 900 °C to 200 °C at 3 °C/s and then cooled slowly to room temperature under vacuum. This led to a final microstructure which contained a small fraction of retained austenite which was used to help validate the reconstruction method.

*Table 1 – Composition of line pipe steel used in this work (in wt%).*

| C | Mn | Cu+Ni+Cr+Mo | V+Nb+Ti |
|---|---|---|---|
| 0.06 | 1.3 | <0.9 | 0.07-0.08 |

EBSD data was collected on a Tescan Amber-X plasma focused ion beam scanning electron microscope (pFIB-SEM) equipped with an Oxford Instruments Symmetry S2 EBSD detector. The EBSD map was collected using a voltage of 20 kV and current of 100 nA for an area of 1 mm$^2$ with 240 nm step size. Patterns were captured at 156 x 128 pixels with an exposure time of 0.16 ms and initially processed online in Oxford Instruments AZtec software, with a minimum of 6 bands required to index the point. A subset measuring 248 x 270 µm is presented in this paper.

The EBSD data was then post-processed offline using Mapsweeper in Oxford Instruments AZtec crystal software [23]. In brief, candidate template diffraction patterns are cross correlated and the best match is selected as the correct phase and crystal orientation [24]. Dynamical templates of ferrite and austenite were generated, and each experimental pattern was compared against both phases. For this work, EBSD-based pattern matching was applied to improve the measurement precision of the ferrite and austenite, and recover more retained austenite (and ferrite) from the unindexed points, and so this processing route was used in Aztec crystal : (1) re-indexing of zero-solutions only (patterns binned to 39 x 32 pixels); orientation refinement (patterns binned to 78 x 66 pixels); repair of isolated map points (patterns binned to 78 x 64 pixels) and a final sweep. Solutions are only accepted if the cross correlation value, R>0.15 and a tolerance of 0.0005. A comparison of the (online) Hough transform vs pattern matched data is shown in Figure 2.

Initial reconstruction of the PAG microstructure used an adapted version of PAG_GUI [25] with an inflation power of 1.4 and an iterative OR using KS as a starting point. PAG grains were identified using a threshold of 5° and a minimum grain size of 100 pixels. Initial reconstructions using points with R > 0.25 only, are presented in Figure 2 for the Hough and Pattern Matched data.

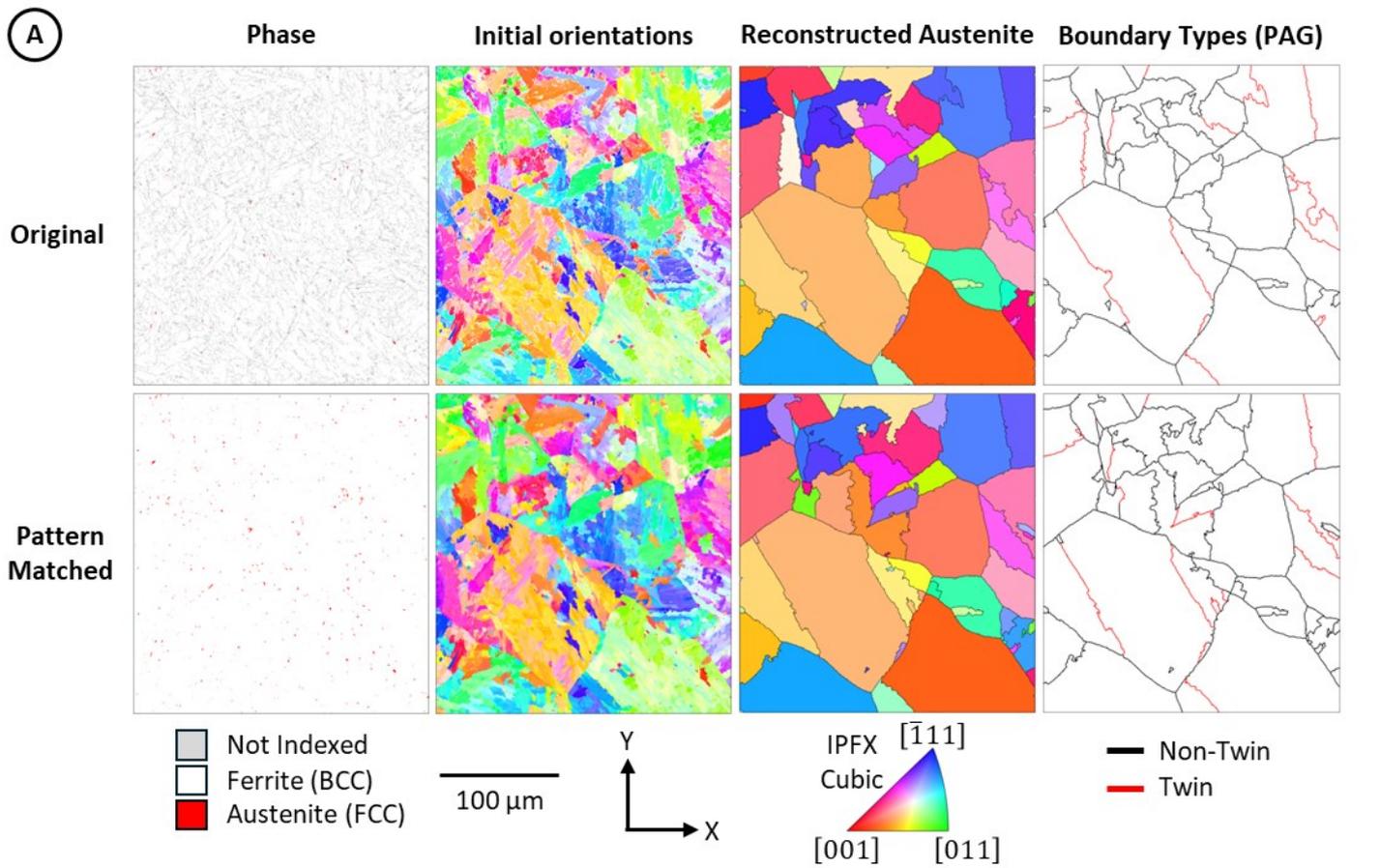

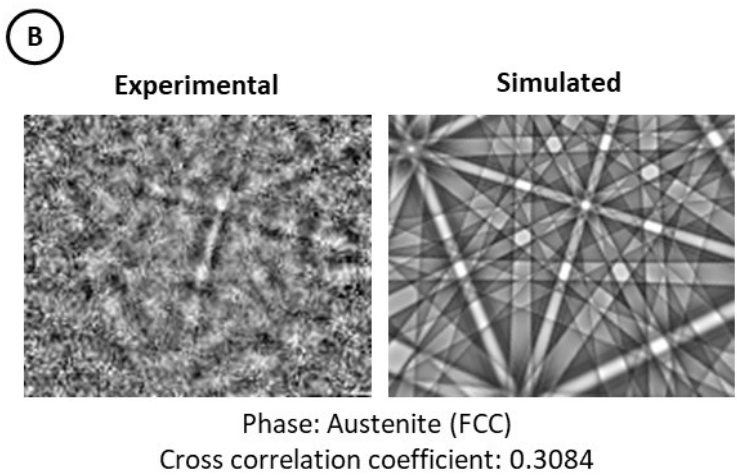

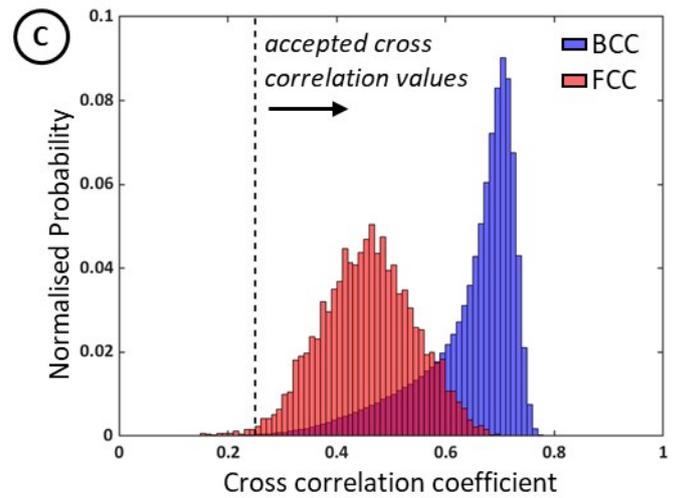

| Phase | Original Hough based Indexing | | After Pattern Matching | |
|---|---|---|---|---|
| | Pixels | % | Pixels | % |
| Not Indexed | 166296 | 14.8 | 1240 | 0.1 |
| Ferrite (BCC) | 956225 | 85.1 | 1117810 | 99.5 |
| Austenite (FCC) | 1175 | 0.1 | 4646 | 0.4 |

*Figure 2 – EBSD map used in this work (A) Comparison of Hough based (original) and pattern matched indexing methods on the indexing and initial reconstruction (using PAG_GUI); (B) An example of an experimental pattern that is unindexed using Hough, but indexed as FCC using pattern matching; (C) Phase statistics with Hough and pattern matched indexing; (D) Cross correlation values for the pattern matching for each phase and the threshold chosen (0.25).*

Figure 2A shows that use of pattern matching significantly changes the location of the austenite-twin grain boundaries (red lines - identified using a tolerance of ±5°) between the reconstructions. In neither case are the twin boundaries straight, and some twin regions appear very small which may indicate uncertainty if not enough variants are present [10].

After this initial reconstruction, the new algorithm (as shown in Figure 3) has been applied, and this is demonstrated for a small sub-region of the dataset. In brief, the algorithm uses a combination of the processed (PAG) grains and the child EBSD orientations to 're-sort' the PAG grain assignments. For each PAG (original and neighbouring), the potential variants are calculated using the mean orientation of the PAG and the experimentally determined OR. This potential variant list is used to test against the observed variants within the neighbour PAG and the potential variants from the initial PAG. If the lowest misorientation angle between the neighbour PAG-associated variant is lower than the misorientation with the assigned PAG, this child variant is 're-sorted' into the neighbour PAG-domain.

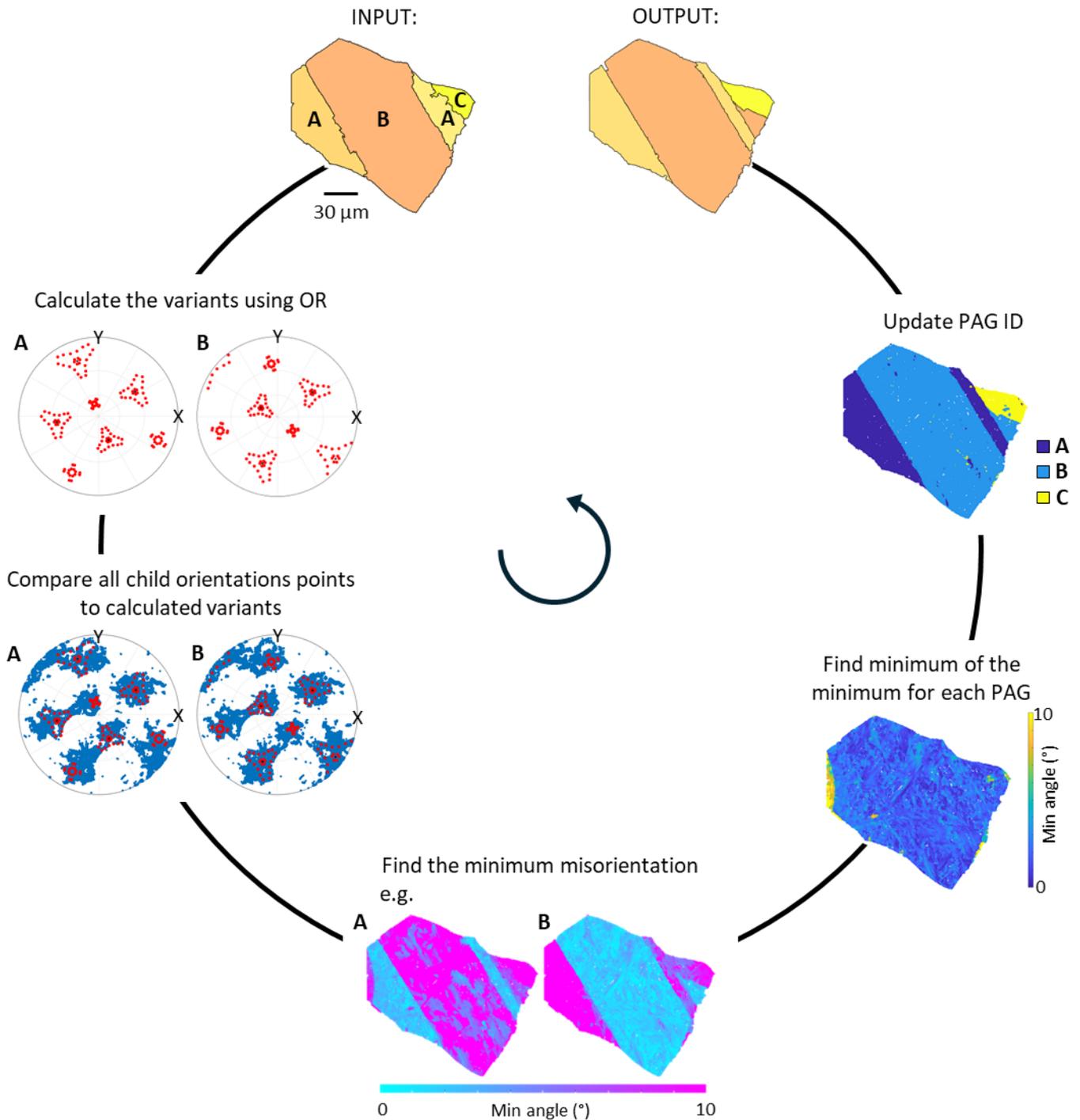

Figure 3 - Re-sort algorithm schematic – A region containing 3 parent grains is used as the input (denoted as A, B and C in the upper left image), with examples for grain A and B shown in the first three steps.

Figure 3 shows that this approach enables a narrow twin, with straight and parallel grain boundary traces, to be reconstructed in the microstructure that is obtained after the re-sort algorithm has been applied. Furthermore, the twin boundary traces within the map have long straight sections, which would be consistent with the coherent twin-interface. In this case, the PAG orientations vary slightly from an ideal twin relationship (misorientation = 58.4°/ ( -12  11  12)) which helps with sorting the variants by providing a slight separation between the variants that are common to both parent orientations.

To validate this method and reconstruction, we compare the outputs (using the example in Figure 3) with and without the re-sort algorithm in Figure 4.

Evaluation of the re-sort algorithm can be performed through comparison of three parameters: (1) the 'fit' (misorientation angle between the closest calculated variant and the child orientation at each point); (2) number of unique variants; (3) and the identified twin grain boundaries. Furthermore, the reconstructed PAG microstructure is also compared with the location and crystal orientations of the retained austenite.

For the fit, a reduction in the fit values should indicate an improvement in the reconstruction, since the calculated variant and experimental variant orientations are now closer.  Next, if there are insufficient and/or a poor combination variants (see [10]) within the region the orientation of the PAG may be not determined correctly, and so regions with a higher number of unique variants being more likely to be correctly identified.  For this microstructure and TMCP processing route, we assume that there are coherent Σ3 twin boundaries in the austenite phase, and this means that we would expect that simple inspection of the microstructure would reveal more linear twin-boundary traces. In the case study shown in Figure 4, we can see that the re-sort algorithm results in an improved PAG reconstruction. Furthermore, in addition to these initial checks, we can use the retained austenite to confirm that the PAG structure is reasonable.

Figure 4 shows initial and re-sorted grain boundary maps with the RA points identified (large dots have been used to make it easier to see locations) and coloured by the misorientation angle between the RA orientation and the PAG orientation assigned at this point. Misorientation angles of <10° show good agreement (green dots), angles of 60 ±5° are likely to indicate the presence of a twin (purple dots), whilst other angles may indicate misidentified PAG regions (orange dots). In this case we see all of the RA points identified as 'twins' in the initial analysis change to well matched after re-sort.

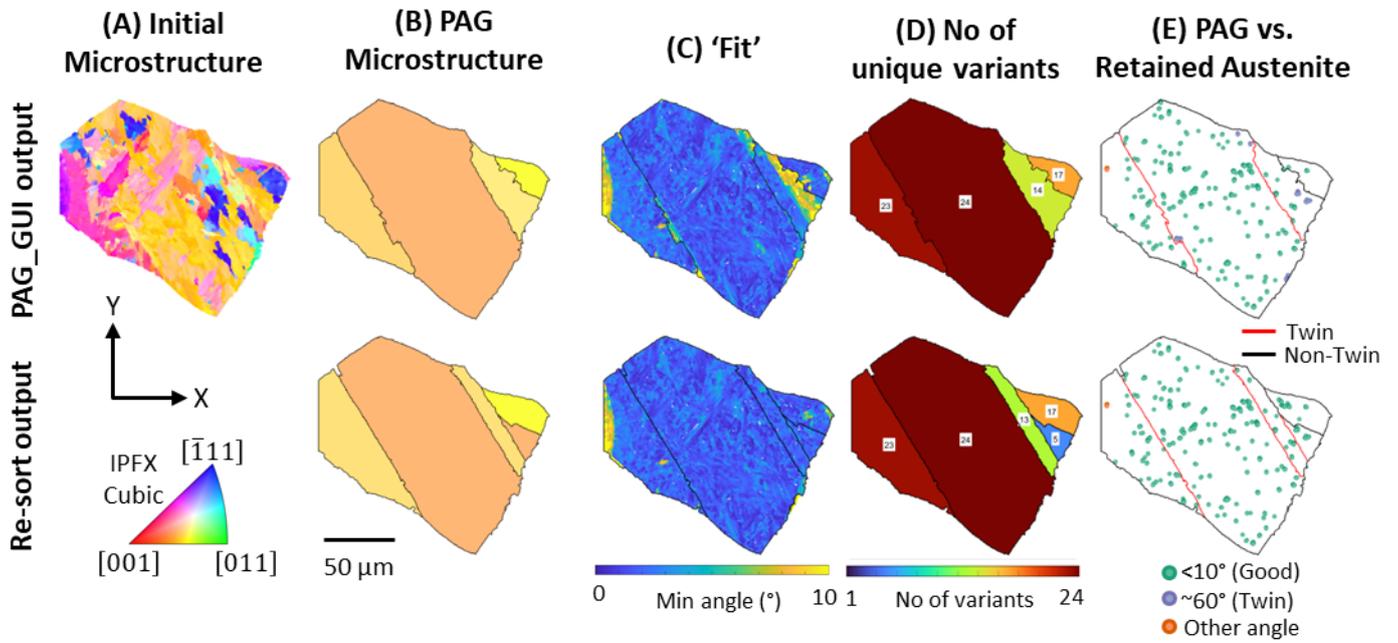

*Figure 4 – Validation of the reconstruction. (A) initial microstructure; PAG_GUI output and re-sort output for (B) austenite microstructure, (C) Fit, (D) No of unique variants per PAG and (E) retained austenite vs. reconstructed orientation and grain boundary network with twin boundaries identified.*

The algorithm has been applied to the entire map, and the results are shown in Figure 5, including quality/validation maps. For the general case, we consider each PAG grain in turn and test all the points using the neighbouring grain orientations in the first instance. After this, PAG grains are re-identified and any that have a low number of variants present (i.e. high uncertainty) are reconsidered to see if they fit well (within a threshold) with a surrounding PAG – this step helps resolve the presence of any island grains that may result due to the common child variants between twins.

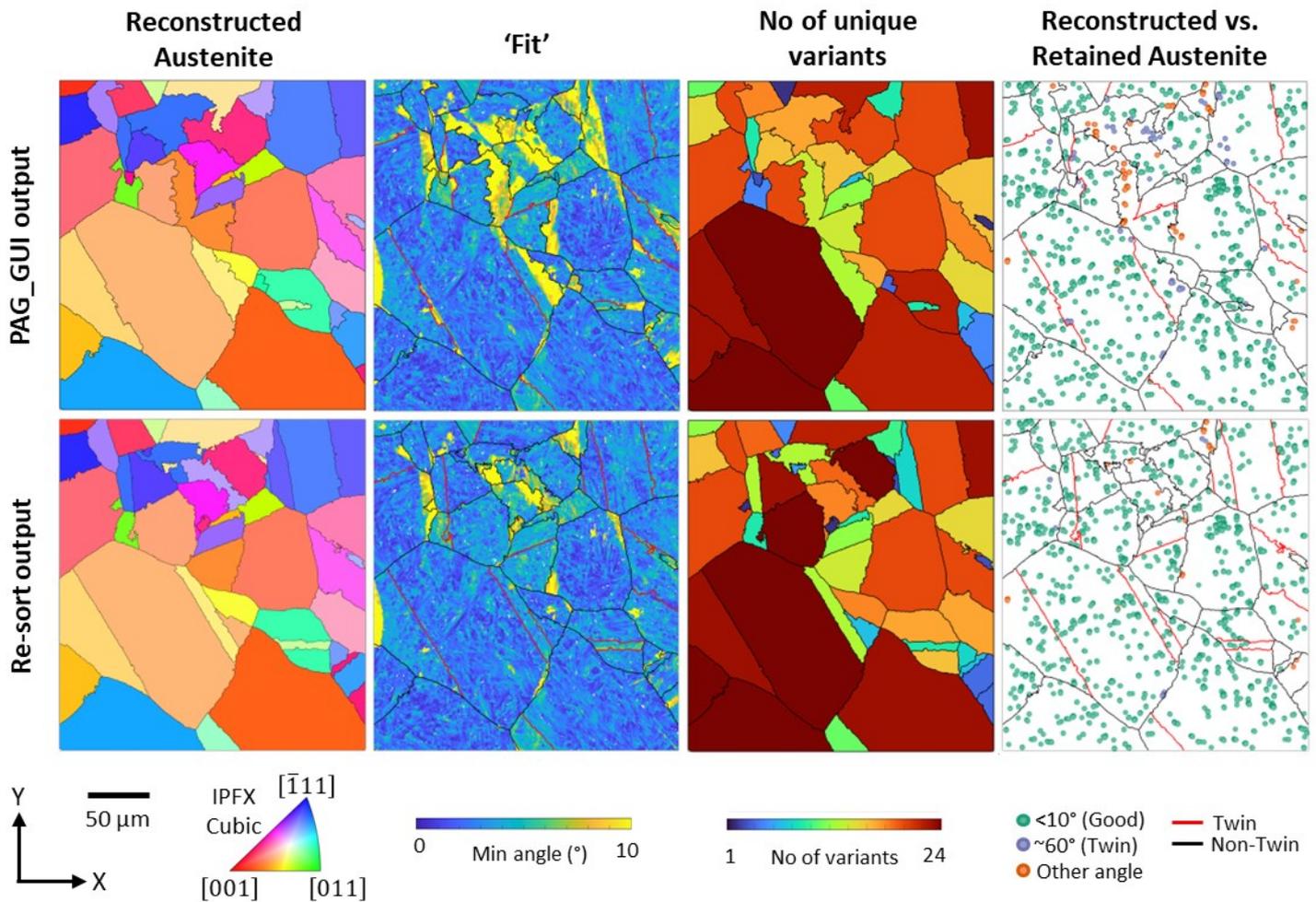

*Figure 5 - Whole map example. PAG_GUI output and re-sort output for (A) austenite microstructure, (B) Fit, (C) No of unique variants per PAG and (D) retained austenite vs. reconstructed orientation and grain boundary network with twin boundaries identified. The initial microstructure is found in Figure 2.*

Figure 5 shows an overall increase in the number of twins identified across the map, with reasonably straight (twin) grain boundary traces identified (no smoothing has been applied). We also see an improvement in the fit values, and generally an increase in the number of unique variants present in each PAG grain/twin grain region. Finally, comparing the retained austenite orientations with the PAG orientations (points within 1 μm of grain boundaries are excluded), the initial PAG reconstruction contains 3353 points (88%) matched, 193 points (5%) twin and 276 (7%) other points; after re-sort there are 3891 (96%) matched, 40 (1%) twin and 108 (3%) other points.

While the algorithm provides significant improvements, there are regions that remain with a poor fit, particularly at the intersections of some of the grains and within some grains. This is most likely due to small PAG domains that were not identified in the original reconstruction: (a) either as the true PAG within this surface map is very small, which results in a significant misorientation of the PAG and retained austenite; (b) or the true PAG is a twin of a region that was not found in the initial reconstruction, and this results in retained austenite that is twinned with respect to the PAG.

In summary, the re-sort algorithm provides a method to address the challenges for PAG reconstruction where there are austenite annealing twins and this has been validated using the retained austenite regions (which were only recovered due to the pattern matching algorithm). We could imagine that this re-sort algorithm could be applied after a variety of different reconstruction methods, provided the OR used is accessible. This algorithm improves the PAG reconstruction significantly, especially for the grain boundary network and correlation with other methods of grain size assessment and development of TMCP steels.

**Acknowledgements**

This work was undertaken in part, thanks to the funding from the Canada Research Chair program (Poole).

The authors would also like to express appreciation and to acknowledge the technical and financial contributions from Evraz NA and NSERC Canada. The Tescan Amber X dual beam microscope used in this study was acquired with financial support from the Canadian Foundation for Innovation (Innovation Fund Project: 39798) and the British Columbia Knowledge Development Fund.

The authors would like to thank Sabyasachi Roy for providing the sample.

**CRediT author statement**

Ruth Birch - conceptualization, investigation, methodology, software, writing - original draft; writing - review & editing; Ben Britton - supervision, methodology, writing - review & editing; Warren Poole - supervision, methodology, writing - review & editing, funding acquisition.